\documentclass[aip,apl,preprint,amsmath,amssymb]{revtex4-1}
\usepackage{graphicx}

\usepackage{dcolumn}
\usepackage{bm}
\usepackage{color}

\newcommand{\jc}{$j_\text{c}$}
\newcommand{\defgl}{\mathrel{\mathop:}=}

\begin{document}



\title[APL]{One-dimensional pinning behavior in Co-doped BaFe$_2$As$_2$ thin films} 


\author{V. Mishev}
\email{vmishev@ati.ac.at.}
\author{W. Seeb\"ock}
\author{M. Eisterer}
\affiliation{Atominstitut, Vienna University of Technology, Stadionallee 2, 1020 Vienna, Austria}

\author{K. Iida}
\author{F. Kurth}
\author{J. H\"anisch}
\author{E. Reich}
\author{B. Holzapfel}
\affiliation{Institute for Metallic Materials, IFW Dresden, 01171 Dresden, Germany}



\begin{abstract}
Angle-resolved transport measurements revealed that planar defects dominate flux pinning in the investigated Co-doped BaFe$_2$As$_2$ thin film. For any given field and temperature, the critical current depends only on the angle between the crystallographic $c$-axis and the applied magnetic field but not on the angle between the current and the field. The critical current is therefore limited only by the in-plane component of the Lorentz force but independent of the out-of-plane component, which is entirely balanced by the pinning force exerted by the planar defects. This one-dimensional pinning behavior shows similarities and differences to intrinsic pinning in layered superconductors. 
\end{abstract}


\maketitle 

Flux pinning is of primary importance for power applications of superconducting materials. The different families of iron-based superconductors offer a rich variety of vortex pinning properties because of their differences in anisotropy and defect structure. Among them, the BaFe$_2$As$_2$ compound (Ba-122) is most promising for applications. Its anisotropy is small and flux pinning can be extremely strong.\cite{Tar12,Fan12} In addition, the limitation of the currents by grain boundaries seems less severe than in the other Fe-based compounds. Furthermore, coated conductors\cite{Iid11} and untextured bulk materials \cite{Wei12} with high critical currents were successfully produced. The material allows for an extraordinarily high density of pinning centers without significant degradation of the transition temperature \cite{Tar12,Fan12} resulting in high critical current densities, \jc, with a weak dependence on the applied magnetic field. The introduction of $c$-axis correlated pinning centers \cite{Tar12,Lee10} leads to a maximum in the angular dependence of the critical current when the field is oriented parallel to the defects and to a reduction in current anisotropy. Here, we report on the opposite limit of a rather clean film with strong pinning only by a few planar defects oriented parallel to the FeAs planes. We performed angle-resolved measurements of the critical current density in a two-axes goniometer to carefully explore the resulting vortex-defect interactions. Similarities to the layered cuprate superconductors\cite{Tac89} were found, although Ba-122 is not expected to form pancake vortices. 

Cobalt-doped Ba-122 films were prepared by pulsed laser deposition using a KrF excimer laser. A 15\,nm thick iron buffer layer was deposited onto a (001) MgO substrate before the 80\,nm thick $10\%$ Co-doped Ba-122 layer was grown.\cite{Iid10b} The transition temperature of this film is 24.7\,K (onset of the resistive transition at zero field). 
	
\begin{figure}
   \centering
	 \includegraphics[width=7cm,clip]{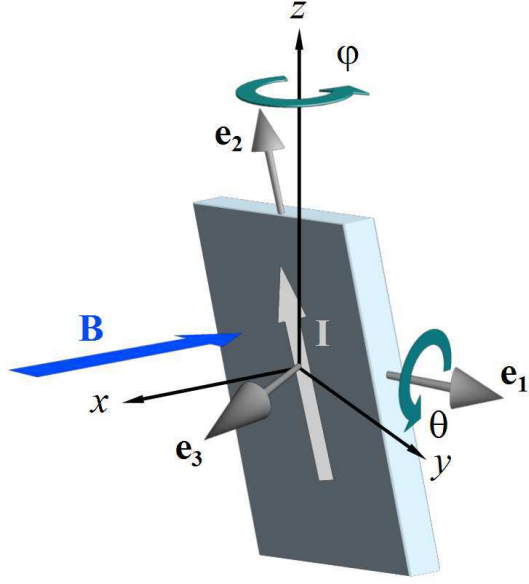}
   \caption{\label{fig:setup} Geometrical representation of the experimental set-up and coordinate
	                            system used for the parametrization of the sample orientation with respect
															to the magnetic field $\bm{B}$.}
\end{figure}

Direct transport measurements were carried out in a helium gas flow cryostat equipped with a $6$\,T split coil magnet. The critical current on a small bridge was evaluated by means of the four-probe method using a criterion of 1\,$\mathrm{\mu}$V/cm. A two-axes goniometer enabled rotations of the sample about two axes which are orthogonal to each other ($z$- and $\bf{e_1}$-axis, Fig.~\ref{fig:setup}). This set-up allows anisotropy measurements of \jc\ under variable Lorentz force (VLF) in addition to the well established measurements under maximum Lorentz force (MLF) resulting in several additional measurement modes such as in-plane scans. In this case, the applied field $H_a$ is always orthogonal to the $c$-axis of the sample and only the orientation of the field with respect to the transport current and, hence, the Lorentz force changes. 

The origin of the angular coordinates $\theta$ and $\varphi$ is chosen such that $H_a$ aligns with the $c$-axis at $\theta = \varphi = 0^\circ$. The measurements included MLF scans ($\theta = 0^\circ$, rotation about $\varphi$), VLF scans ($\varphi =$ const., rotation about $\theta$) and in-plane scans ($\theta = 90^\circ$, rotation about $\varphi$). The temperature range comprised seven different temperatures ranging between $20$\,K and $5$\,K.

With the azimuthal and polar angle, $\varphi$ and $\theta$, one can easily express a parametrization of the unit vectors corresponding to the three crystallographic directions. Note that the $xyz$-coordinate system as well as the magnetic field vector $\bm{B}$ are fixed within the cryostat. According to Fig.~\ref{fig:setup}, the unit vectors are then given by
\begin{eqnarray}
{\bf e_1} =
\left(\!\!\begin{array}{c}
	-\sin\varphi \\
	 \phantom{-}\cos\varphi \\
	    0      \end{array}\right),\;
{\bf e_2} =
\left(\begin{array}{c}
	 \cos\varphi \sin\theta \\
	 \sin\varphi \sin\theta \\
	 \cos\theta           \end{array}\right),\;
{\bf e_3} =
\left(\begin{array}{c}
	\cos\varphi \cos\theta \\
	\sin\varphi \cos\theta \\
	-\sin\theta     			\end{array}\right).
\label{eq:unit_vectors}
\end{eqnarray}
Since the current through the sample is supposed to flow along the $\bf{e_2}$-axis, the resulting Lorentz force (here and in the following the Lorentz force has the physical dimension of a force per length) becomes
\begin{equation}
	{\bm F_\text{L}} = I_c B
	\left(\begin{array}{c}
				  0                   \\
					\cos\theta          \\
				- \sin\varphi \sin\theta   \end{array}\right) ,
	\label{eq:LF}
\end{equation}
where $I_\text{c}$ and $B$ are the absolute values of the critical current and the magnetic field, respectively. Hence, the projection of ${\bm F_\text{L}}$ onto the $ab$-plane of the sample is
\begin{equation}
	{\bm F^\|_\text{L}} = -I_\text{c} B \cos\alpha
	\left(\!\!\begin{array}{c}
				-\sin\varphi \\
				\phantom{-} \cos\varphi \\
				   0
			\end{array}\right) ,
	\label{eq:inplaneLFalpha}
\end{equation}                
where $\cos\alpha\!\defgl\!\cos\varphi \cos\theta$ and $\alpha$ denotes the angle between the magnetic field and the crystallographic $c$-axis of the sample. The absolute value of the in-plane component of the Lorentz force is, therefore, only a function of $\alpha$:
\begin{equation}
	F^{\|}_\text{L} = I_\text{c} B \cos\alpha \,.
	\label{eq:absFalpha}
\end{equation}
\begin{figure}
   \includegraphics{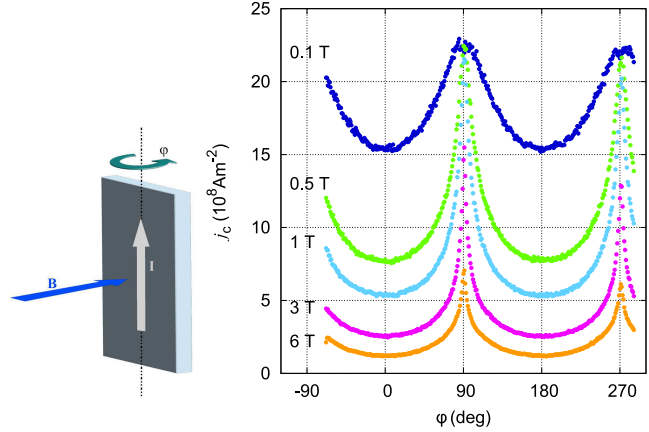} 
	 \caption{Angle-resolved measurements of $j_c(\varphi,\theta=0)$ at $T=10$\,K for
	          various fields. The $j_c$-curves show a pronounced $ab$-peak due to planar defects.}
	 \label{fig:jc_phi_10K}
\end{figure}
\begin{figure}
  \includegraphics{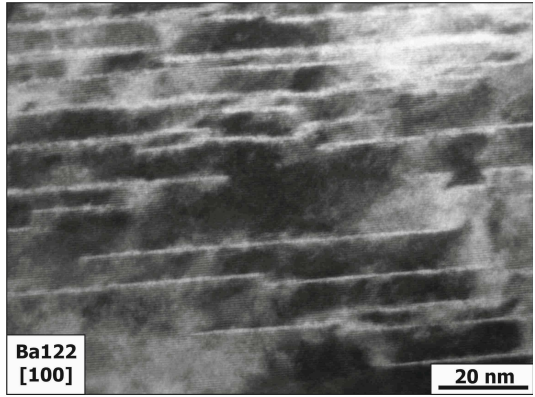}
	\caption{\label{fig:TEM_image} TEM bright field image of another Ba-122 film prepared under the same conditions as the film
	                investigated in this letter. Planar defects parallel to the $ab$-planes are clearly visible.}
\end{figure}
Fig.~\ref{fig:jc_phi_10K} shows the \jc\-anisotropy in standard MLF configuration at 10\,K for various fields. No $c$-axis peak is observed, and the critical current exhibits a sharp peak when the field is oriented parallel to the $ab$-planes. Since the $c$-axis coherence length in Ba-122 ($\xi_c\approx 1.2$\,nm\cite{Yam09}) is similar to the distance between the FeAs layers (smaller than the lattice parameter $c \approx 1.3$\,nm\cite{Rot08}), the condensation energy is not expected to be suppressed significantly between the layers and intrinsic pinning can be ruled out as a reason for the $ab$-peak. Instead, planar defects seem responsible, which were revealed by Bright field transmission electron microscopy (TEM) in an equally prepared, although slightly thicker film (Fig.~\ref{fig:TEM_image}).

\begin{figure}
   \includegraphics{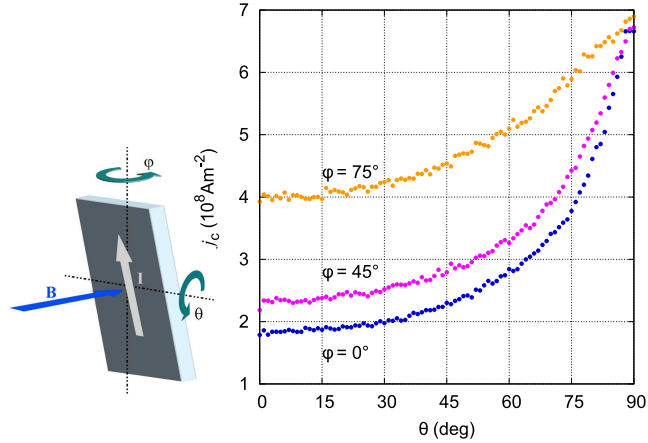}  
	 \caption{Variable Lorentz force (VLF) scans $j_\text{c}(\varphi,\theta)$ at
	         $T=15$\,K and $B=1$\,T.}
	 \label{fig:VLF_1T_15K}
\end{figure}

\begin{figure}
   \includegraphics{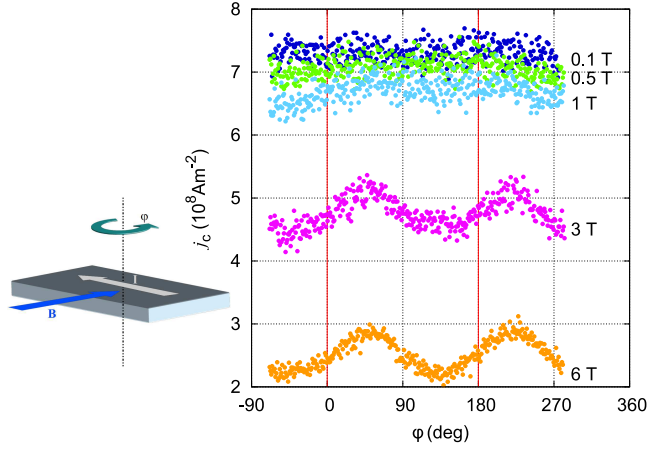}  
	 \caption{In-plane scans of $j_\text{c}(\varphi,\theta=90^\circ)$ at $T=15$\,K for various fields.
	          The out-of-plane component of the Lorentz force becomes zero at $0^\circ$ and $180^\circ$.}
	 \label{fig:in_plane_15K}
\end{figure}

To investigate the critical current behavior under variable Lorentz force, the angle $\theta$ was varied at several fixed angles $\varphi$. Fig.~\ref{fig:VLF_1T_15K} shows typical results at 15\,K and 1\,T. Note the fact that all curves converge at $\theta=90^\circ$, independent of the angle $\varphi$.
A special case of the VLF configuration occurs for $\theta=90^\circ$ and variable $\varphi$, where the applied magnetic field and the current remain in the same plane, i.e.\!\! parallel to the crystallographic $ab$-planes. The resulting Lorentz force varies between zero and its maximum and is always perpendicular to the $ab$-planes and, in the present case, also to the planar defects. Typical so called in-plane scans are shown in Fig.~\ref{fig:in_plane_15K} for various fields at 15\,K. At low fields, hardly any variation in \jc{} is detectable -- apart from the scatter of the data --, which means that the out-of-plane component of the Lorentz force does not have any influence, while depinning is triggered by the in-plane component resulting from the self-field of the current. Since \jc\ does not change with angle, the self-field and the in-plane component of the Lorentz force remain constant, as required for this explanation. Applying higher fields results in a significant variation with $\varphi$ which cannot be considered as a consequence of the changing Lorentz force because the minimima and maxima of \jc{} do not coincide with those of the Lorentz force at 0 and $90^\circ$, respectively. We ascribe the variation in \jc{} to geometric imperfections. At $6$\,T \jc{} changes by about $\pm 15\%$ which corresponds to a tilt angle between the field and the $ab$-plane of about $1.5^\circ$ only, since the $ab$-peak is very sharp under these conditions (c.f. Fig.~\ref{fig:jc_phi_10K}). A small tilt of the $ab$-planes with respect to the sample holder together with a slight misalignment of the rotation axis would explain the observed behavior.
\begin{figure}
  \includegraphics{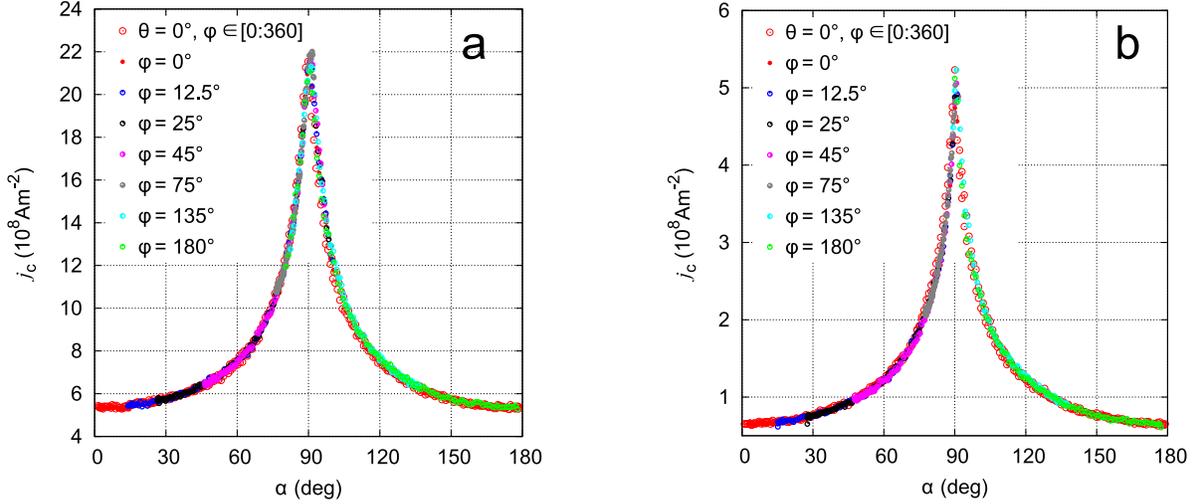}
	\caption{\label{fig:master_15_10}  $j_\text{c}(\alpha)$ at $T=10$\,K, $B = 1$\,T (a) and at $T=15$\,K,
	                $B = 3$\,T (b). Any combination of	$\varphi$ and $\theta$ can be scaled to one single
									angle $\alpha$, which denotes the angle between the crystallographic $c$-axis and
									the magnetic field.}
\end{figure}

The independence of the critical current on the out-of-plane component of the Lorentz force is not only observed in the in-plane scans ($\theta=90^\circ$), but all data of $j_\text{c}(\varphi,\theta\!=\!const)$ and $j_\text{c}(\theta,\varphi\!=\!const)$ collapse at each field and temperature when plotted as a function of $\alpha$. Since $\alpha$ denotes the angle between the crystallographic $c$-axis and the magnetic field, this procedure corresponds to mapping all these data to the MLF configuration. Reducing the out-of-plane component of the Lorentz force -- which is not a unique function of $\alpha$ -- does not change \jc{}. Thus, the critical current is limited by the in-plane component of ${\bm F_\text{L}}$ only. Fig.~\ref{fig:master_15_10} shows exemplary results for $T=15$\,K, $B=3$\,T and $T=10$\,K, $B=1$\,T, respectively. It can be concluded that flux pinning is extremely anisotropic in these films and that pinning on the planar defects by far exceeds any random or $c$-axis correlated contributions. However, the planar defects can balance only one component of the Lorentz force, while the other one limits \jc{} leading to a one-dimensional pinning behavior. (In reality, the projection space of the Lorentz force onto the $ab$-planes is two-dimensional, but the rotational symmetry reduces its formal dimensionality.)

The situation resembles the intrinsic pinning of pancake vortices in layered super\-con\-duc\-tors, which would also result in the observed scaling with $\alpha$. However, our data do not follow the angular dependence predicted by the Tachiki-Takahashi model.\cite{Tac89} Instead, at low fields (e.g. up to about 1\,T at 10\,K), where the vortex-vortex interaction is small, data obtained at different magnetic fields coalesce when plotted directly as a function of the out-of-plane component of the magnetic field (i.e. $B\cos\alpha$), as expected from the current limiting mechanism. This difference to the Tachiki-Takahashi behavior is not unexpected since the ratio between coherence length and interlayer spacing is incompatible with the formation of pancake vortices, on which that model is based. At higher fields data obtained at different fields cannot be scaled by the Tachiki-Takahashi approach nor by $B\cos\alpha$, which is a consequence of the vortex-vortex interactions. This highlights the advantage of the $\alpha$ scaling (data obtained at VLF) for the determination of pinning dimensionality compared to scaling approaches for changing magnetic field because it is free of any assumptions of the vortex properties (changing vortex density).  

In conclusion, a careful analysis of the \jc{} anisotropy revealed a one-dimensional pinning behavior in Co-doped Ba-122 films containing correlated planar defects. Only the component of the Lorentz force parallel to the defects is responsible for the critical current density in the films, and for any given field and temperature the angular dependence of \jc{} is a unique function of the angle between the magnetic field and the crystallographic $c$-axis only.

\begin{acknowledgments}
We acknowledge Bernhard Berger for his contributions to the measurements and Prof. H. W. Weber for fruitful discussions. This work was supported by the Austrian Science Fund (FWF): P22837-N20 and by the European-Japanese collaborative project SUPER-IRON (No. 283204).
\end{acknowledgments}

\end{document}